\documentclass[preprint,5p,twocolumn]{elsarticle}

\usepackage{amssymb}
\usepackage{subfigure}
\usepackage{multicol}
\usepackage{graphicx}
\biboptions{longnamesfirst,square,sort&compress,comma}
\journal{Physics Letters A}

\def\figref#1{\mbox{Figure~\ref{#1}}}
\def\figrefs#1{\mbox{Figures~\ref{#1}}}
\def\Figref#1{\mbox{Fig.~\ref{#1}}}
\def\Figrefs#1{\mbox{Figs.~\ref{#1}}}
\def\FIGref#1{\mbox{\ref{#1}}}
\def\Eqref#1{\mbox{Eq.~(\ref{#1})}}

\def\eqref#1{\mbox{(\ref{#1})}}
\def\secref#1{\mbox{Sec.~\ref{#1}}}

\def\tabref#1{\mbox{Tab.~\ref{#1}}}
\begin{document}

\begin{frontmatter}
\title{Counter operation in nonlinear micro-electro-mechanical resonators}

\author{Atsushi Yao and Takashi Hikihara \vspace{5mm}
\\ {\em Department of Electrical Engineering, Kyoto University, \\Katsura, Nishikyo, Kyoto 615-8510 Japan} \vspace{3mm}
\\ yao@dove.kuee.kyoto-u.ac.jp (A.~Yao), hikihara.takashi.2n@kyoto-u.ac.jp (T.~Hikihara)}

\begin{abstract}
This paper discusses a logical operation of multi-memories that 
consist of coupled nonlinear micro-electro-mechanical systems (MEMS) resonators. 
A MEMS resonator shows two coexisting stable states when nonlinear responses appear. 
Previous studies addressed that a micro- or nano-electrical-mechanical resonator can be utilized as a mechanical 1-bit memory or mechanical logic gates. 
The next phase is the development of logic system with coupled multi-resonators. 
From the viewpoint of application of nonlinear dynamics in coupled MEMS resonators, 
we show the first experimental success of the controlling nonlinear behavior as a 2-bit binary counter. 
\end{abstract}

\end{frontmatter}

\section{Introduction}
A micro-electro-mechanical resonator is a kind of devices fabricated using micro-electro-mechanical systems (MEMS) technology. 
A MEMS resonator substantially shows nonlinear responses at large excitation force. 
The dynamics of a nonlinear micro- or nano-electro-mechanical resonator 
is described by the Duffing equation~\cite{Badzey2004, Mestrom2008, Unterreithmeier2010b, Naik2012, antonio2012frequency, Atsushi2012}. 
At nonlinear responses, an amplitude-frequency response curve bends toward higher or lower frequencies due to 
a hard or soft spring effect~\cite{JohnA.Pelesko2003}. 
In particular, at a fixed excitation frequency in the hysteresis region, 
a nonlinear MEMS resonator has 
two stable periodic states and an unstable periodic state~\cite{V.Kaajakari2009}. 

Recently, many studies focus on mechanical computation 
especially in micro- 
and nano-electro-mechanical resonators~\cite{Badzey2004, Mahboob2008, Noh2010, Unterreithmeier2010b, Yao2012, Atsushi2012, uranga2013exploitation, Masmanidis2007, guerra2010noise, Mahboob2011, hatanaka2012electromechanical}. 
A mechanical memory device is a prospective application of 
nonlinear micro- or nano-electro-mechanical resonators~\cite{Badzey2004, Mahboob2008, Noh2010, Unterreithmeier2010b, Yao2012, Atsushi2012, uranga2013exploitation}. 
Previous studies demonstrated that a micro- or nano-electro-mechanical resonator 
can be applied to execute mechanical logic gates~\cite{Masmanidis2007, guerra2010noise, Mahboob2011, hatanaka2012electromechanical}. 
It is natural to expect that the next phase is the development of logic system with coupled multi-resonators. 

This paper discusses a 2-bit binary counter~\cite{Morris1991} from the viewpoint of application of nonlinear dynamics in coupled MEMS memories. 
The authors experimentally showed the switching control between two coexisting stable states 
at a fixed excitation frequency in a nonlinear MEMS resonator~\cite{Yao2012}. 
Based on our previous study, here we show the first experimental success of 
the controlling nonlinear behavior in two coupled MEMS resonators as a 2-bit binary counter. 

The overview of this paper is organized as follows. \secref{section2} presents two fabricated MEMS resonators and its four coexisting stable states. 
\secref{section3} explains the switching control system in the coupled nonlinear MEMS resonators. 
In \secref{section4}, the switching control sequence of counter operation is experimentally achieved and examined.

\section{MEMS resonator and its coexisting stable states}\label{section2}
\begin{figure}[!t]
 \begin{center}
 \includegraphics[width=1\linewidth]{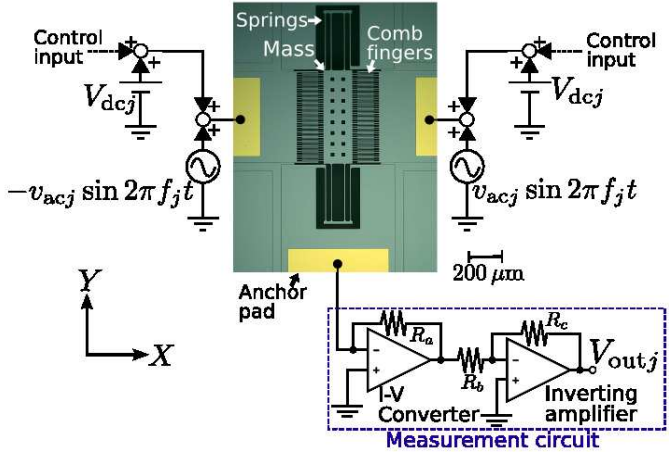} 
  \caption{Schematic diagram of single comb-drive resonator and measurement system. Fabricated resonator has a perforated mass with length, width and thickness of $575$, $175$, and $25$\,$\mu$m. The perforated mass is supported by folded beams, which are used as springs. They are connected to anchors. The MEMS resonator has two comb capacitors. Resistors $R_{a}$, $R_{b}$, and $R_{c}$ are set at 1\,M$\Omega$, 1\,k$\Omega$, and 100\,k$\Omega$. Output voltage $V_{{\rm out}j}$ depends on the amplitude and the phase of the displacement.}
  \vspace{0mm}
\label{fig1}
 \end{center}
\end{figure}

\figref{fig1} shows an electrostatically driven comb-drive resonator that was fabricated with silicon on insulator (SOI) 
technology~\cite{Naik2012, Naik2011, Naik2011a}. 
When a MEMS resonator is actuated by applying an ac excitation voltage with a dc bias voltage 
between the mass and the electrodes, the mass vibrates in the $X$--direction with a weak link to the $Y$--direction. 
In this paper, we utilize two comb-drive resonators. 

\figref{fig1} also presents a schematic diagram of the experimental setup in the differential measurement~\cite{V.Kempe2011} 
for a single MEMS resonator as described in Ref.~\cite{Yao2012}. 
In the differential measurement for a single MEMS resonator, excitation force $F_{j}$ and output voltage $V_{{\rm out}j}$ 
are obtained by the following equations: 
\begin{eqnarray}
F_{j} &=&  4  \varepsilon N \frac{h}{d} V_{{\rm dc}j} v_{{\rm ac}j}\sin 2 \pi f_{j} t,  \label{Ffirst} \\
V_{{\rm out}j} &=& 8\times 10^{8} \pi f_{j}  \varepsilon N  \frac{h}{d} v_{{\rm ac}j} A_{j}\sin (4 \pi f_{j} t + \phi_{j}),  \label{v}   
\end{eqnarray}
where $j = 1,2$. 
Here $f_{j}$ denotes the excitation frequency of the $j$-th MEMS resonator in the coupled system, 
$A_{j}$ the amplitude of the displacement, and $\phi_{j}$ the phase. 
$ N~(=39)$ denotes the comb number of each resonator, $h ~(=25 $ $\mu$m$)$ the finger height, 
$d ~(=3 $\,$\mu$m$)$ the gap between the fingers, and $\varepsilon ~(= 8.85 \times 10^{-12}$\,F/m$)$ the permittivity. 
We set the dc bias voltage $V_{{\rm dc}j}$, the ac excitation amplitude $v_{{\rm ac}j}$, and the pressure 
at $-0.15$\,V, $0.6$\,V, and around $12$\,Pa at room temperature. 
The first (second) MEMS resonator is called Res.~1 (Res.~2) from here on. 

\begin{figure}[!tb]
 \begin{center}
  \begin{minipage}{0.48\hsize}
   \subfigure[Amplitude-frequency response curves of Res.~1.]{
    \centering
     \includegraphics[width=0.97\linewidth]{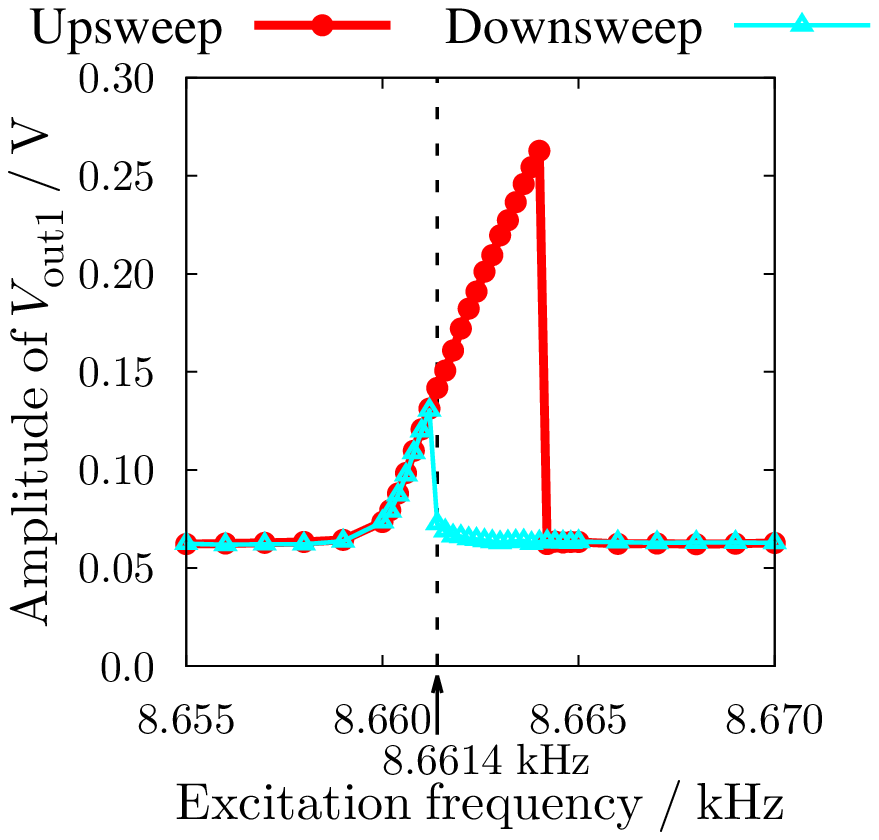}
   \label{Vout1}}
  \end{minipage}
 \hspace{1mm}
  \begin{minipage}{0.48\hsize}
   \subfigure[Two coexisting stable states at $f_{1}=8.6614$\,kHz of Res.~1.]{
    \centering
    \includegraphics[width=.98\linewidth]{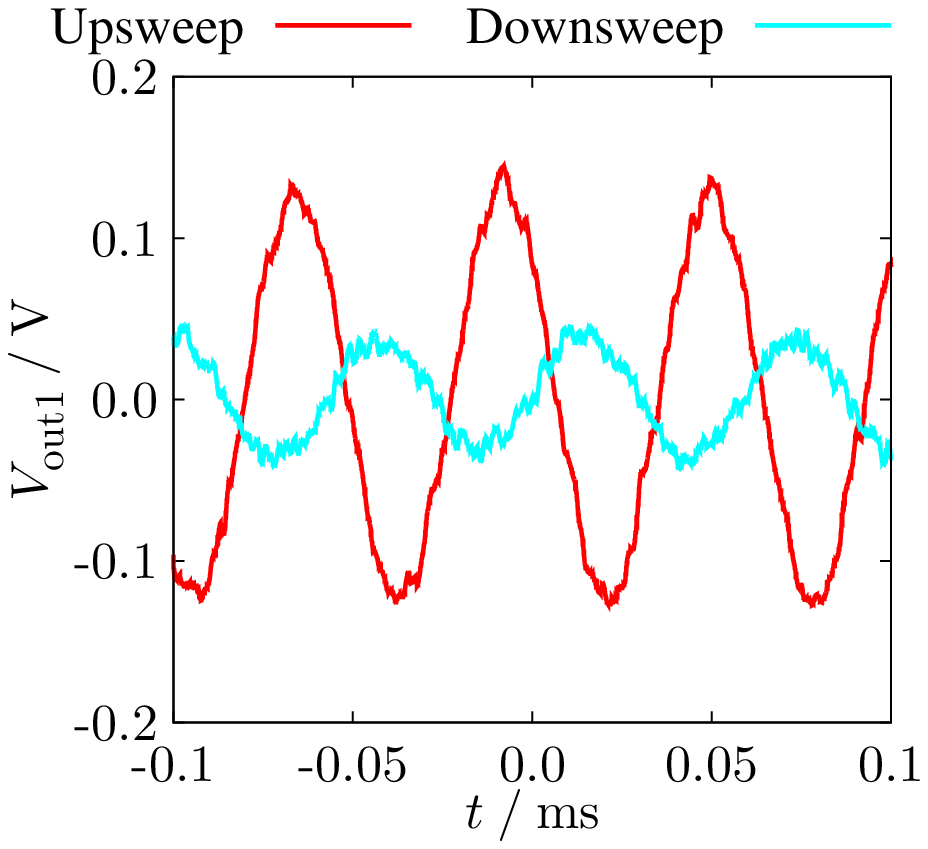}
   \label{UpandDown_Time1}}
   \end{minipage}
\vspace{0mm}
 \caption{Displacement measurement by differential configuration in Res.~1. At any given frequency in the hysteretic regime, the resonator can exist in two distinct amplitude states.} 
\vspace{0mm}
\label{MeasurementDate1}
  \end{center}
\end{figure}

\begin{figure}[!tb]
 \begin{center}
  \begin{minipage}{0.48\hsize}
   \subfigure[Amplitude-frequency response curves of Res.~2.]{
     \centering
     \includegraphics[width=.97\linewidth]{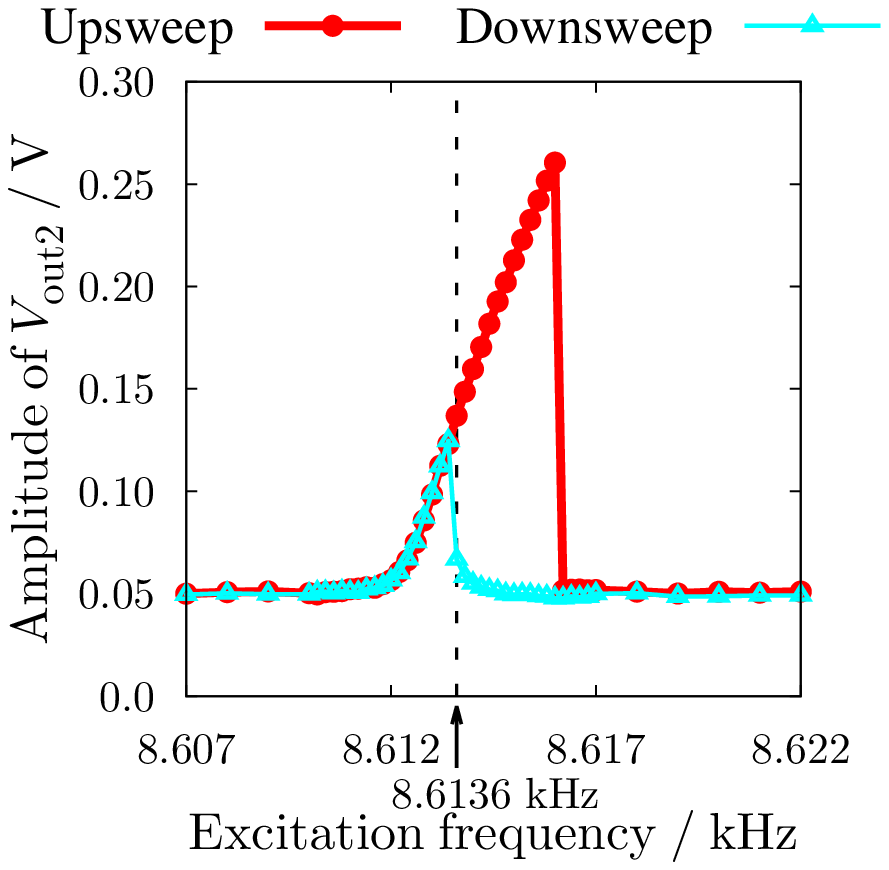}
   \label{Vout2}}
  \end{minipage}
  \hspace{1mm}
  \begin{minipage}{0.48\hsize}
   \subfigure[Two coexisting stable states at $f_{2}=8.6136$\,kHz of Res.~2.]{
    \centering
    \includegraphics[width=.98\linewidth]{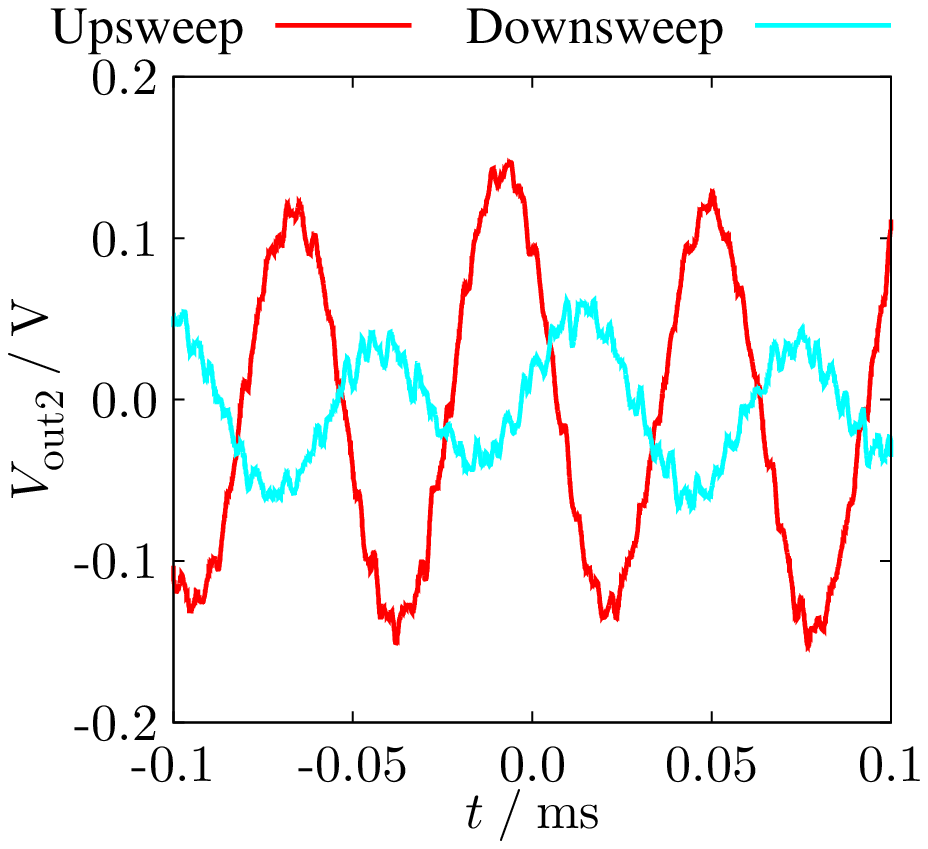}
   \label{UpandDown_Time2}}
   \end{minipage}
\vspace{0mm}
 \caption{Displacement measurement by differential configuration in Res.~2. Two different vibrational states coexist.}
\vspace{0mm}
\label{MeasurementDate2}
  \end{center}
\end{figure}

\figref{Vout1} (\FIGref{Vout2}) shows the experimentally obtained frequency response curves in Res.~1 (Res.~2). 
The red and aqua lines correspond to the responses at the upsweep and the downsweep of frequency, respectively. 
The frequency response curves strongly depend on the sweep direction in the hysteresis region. 
We found that the behavior exhibited by the MEMS resonator qualitatively resemble to each other. 
Two stable states coexist at $8.6612$\,kHz $< f_{1} < 8.6642$\,kHz in \Figref{Vout1} and at  $8.6134$\,kHz $< f_{2} < 8.6162$\,kHz 
in \Figref{Vout2}. 
It was considered that the difference of hysteresis is caused by different 
doping angle, debris deposited during fabrication and die separation, and/or minute cracks~\cite{Naik2011}. 

\figrefs{UpandDown_Time1} and \FIGref{UpandDown_Time2} show the oscillogram of two stable periodic vibrations at $8.6614$\,kHz in Res.~1 
and at $8.6136$\,kHz in Res.~2. 
The red and aqua lines are averaged out over $32$ measurements. 
Below, the excitation frequency is fixed at $8.6614$\,kHz in Res.~1 and at $8.6136$\,kHz in Res.~2. 
The large (small) amplitude vibration is regarded as the ``$1$'' (``$0$'') state for each resonator. 
In addition, Res.~1 holds the first bit and Res.~2 the second bit.

\section{Switching control method}\label{section3}
\begin{figure}[!t]
 \centering
  \includegraphics[width=1\linewidth]{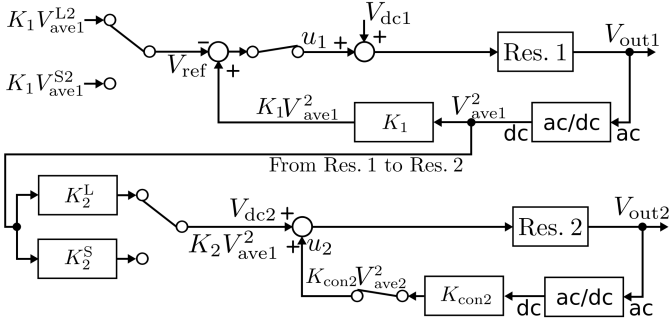}
 \caption{Switching control system.}
 \label{1st}
\end{figure}

\figref{1st} shows the switching control system. 
A binary counter is a sequential system that goes through a 
prescribed sequence of states upon the application of clock 
signals~\cite{Morris1991}. 
In the binary counter, the output transition of one MEMS resonator triggers the switching control of other MEMS resonators. 
In order to realize a 2-bit binary counter, two MEMS resonators are interconnected in a unidirectional coupled system. 
Although the electrical noise here appears due to electrical coupling between two MEMS resonators, 
there does not happen any fault of switching control. 

Based on our previous work~\cite{Yao2012}, we construct a switching control system with the feedback control in Res.~1 
and apply the control input as a slowly changing dc voltage to the MEMS resonator. 
The slowly changing dc voltage is given as a square average dc voltage, 
to which the output voltage is converted by an analog multiplier and a low pass filter of the operational amplifier. 
In Res.~1, excitation force $F_{1}$ under control and control input $u_{1}$ are defined by the following equations: 
\begin{eqnarray}
F_{1} &=&  4  \varepsilon N \frac{h}{d} (V_{{\rm dc}1} + u_{\rm 1}) v_{{\rm ac}1}\sin 2 \pi f_{1} t, \label{fcontrol1}\\
u_{1} &=& - V_{\rm ref} + K_{1} V_{\rm ave1}^{2},   \label{udc1}
\end{eqnarray}
where $K_{\rm 1}~(=9.1)$ denotes the feedback gain, $V_{\rm ref}$ the external reference signal, and $V_{{\rm ave}1}^{2}$ 
the square average dc voltage of Res.~1. 
The external reference signal $V_{\rm ref}$ is set at $K_{1} V_{\rm ave1}^{{\rm L}2}$ ($K_{1} V_{\rm ave1}^{{\rm S}2}$) 
when the state is requested to switch to a large (small) amplitude vibration. 
Here, $V^{{\rm L}2}_{\rm ave1}$ ($V^{{\rm S}2}_{\rm ave1}$) corresponds to the targeted square average dc voltage 
for the large (small) amplitude vibration. 
The MEMS resonator exhibits the hysteretic behavior with respect to the excitation force 
at a fixed excitation frequency~\cite{guerra2010noise, uranga2013exploitation}. 
We need to decrease (increase) the amplitude of the excitation force under control 
when the state is switched to the small (large) amplitude vibration.
 
The 2-bit binary counter consists of a series connection of two MEMS resonators, 
where the output of Res.~1 is connected to the input of Res.~2. 
As shown in \Eqref{fcontrol1}, the excitation force depends on both dc bias voltage $V_{\rm dc1}$ and control input $u_{1}$. 
Here, we propose that the dc bias voltage $V_{{\rm dc}2}$ of Res.~2 is given as the input 
depending on the square average dc voltage $V_{\rm ave1}^{2}$ of Res.~1. 
In Res.~2, excitation force $F_{2}$ under control, dc bias voltage $V_{{\rm dc}2}$, and 
control input $u_{2}$ are obtained by the following equations:  
\begin{eqnarray}
F_{2} &=&  4  \varepsilon N \frac{h}{d} (V_{\rm dc2} + u_{\rm 2}) v_{\rm ac2}\sin 2 \pi f_{2} t, \label{fcontrol2}\\
V_{{\rm dc}2} &=&  K_{\rm 2}  V_{\rm ave1}^{2},  \label{udc2}\\
u_{2} &=&  K_{\rm con2}  V_{\rm ave2}^{2},  \label{Vdc2}
\end{eqnarray}
where $K_{\rm 2}$ denotes the gain, $K_{\rm con2}~(= -3.9)$ the control gain, and $V_{\rm ave2}^{2}$ the square average dc voltage of Res.~2. 
When the state of Res.~1 is switched to the large (small) amplitude vibration, the gain $K_{2}$ is set at $K^{\rm L}_{2}$ ($K^{\rm S}_{2}$). 
Here, $K^{\rm L}_{2} V_{\rm ave1}^{{\rm L}2} $ and $K^{\rm S}_{2} V_{\rm ave1}^{{\rm S}2} $ are adjusted at $-0.15$\,V. 
Furthermore, the feedback gain $K_{1}$ and the control gain $K_{\rm con2}$ are swept within the operating range and are adjusted at $9.1$ and $-3.9$. 

\begin{table}[!b]
  \caption{Count sequence for 2-bit binary counter.}
  \label{count}
  \begin{center}
 \begin{tabular}{c|cc} 
 Clock & Res.~2 & Res.~1  \\
  & $V_{\rm out2}$ & $V_{\rm out1}$\\ \hline
 0 & ``0'' & ``0'' \\
 1 & ``0'' & ``1'' \\
 2 & ``1'' & ``0'' \\
 3 & ``1'' & ``1'' \\ \hline
  \end{tabular}
  \end{center}
\end{table}

\section{Switching control results and discussion}\label{section4}

\begin{figure}[!b]
 \begin{center}
 \hspace{-1mm}  \begin{minipage}{0.47\hsize}
   \subfigure[Switching control from ``0'' to ``1'' in Res.~1 at $V_{\rm ref} = K_{1} V_{\rm ave1}^{{\rm L}2}$.]{
    \centering
    \includegraphics[width=0.99\linewidth]{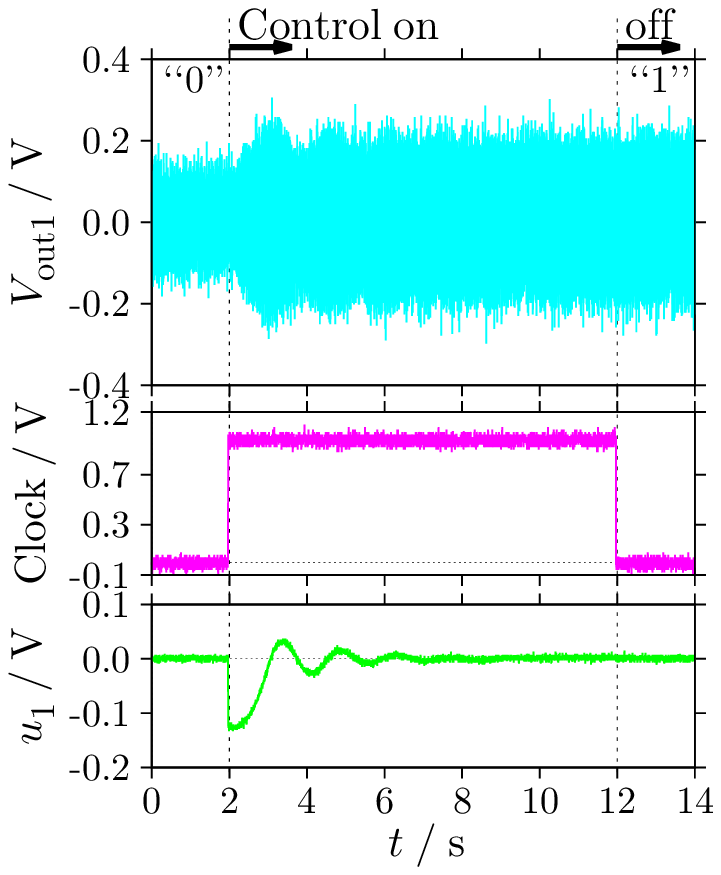}
      \label{00to01Res1}}
  \end{minipage}
  \hspace{1.5mm}
  \begin{minipage}{0.47\hsize}
   \subfigure[Switching control from ``0'' to ``0'' in Res.~2 at $K_{2}=K_{2}^{\rm L}$.]{
    \centering
    \includegraphics[width=0.99\linewidth]{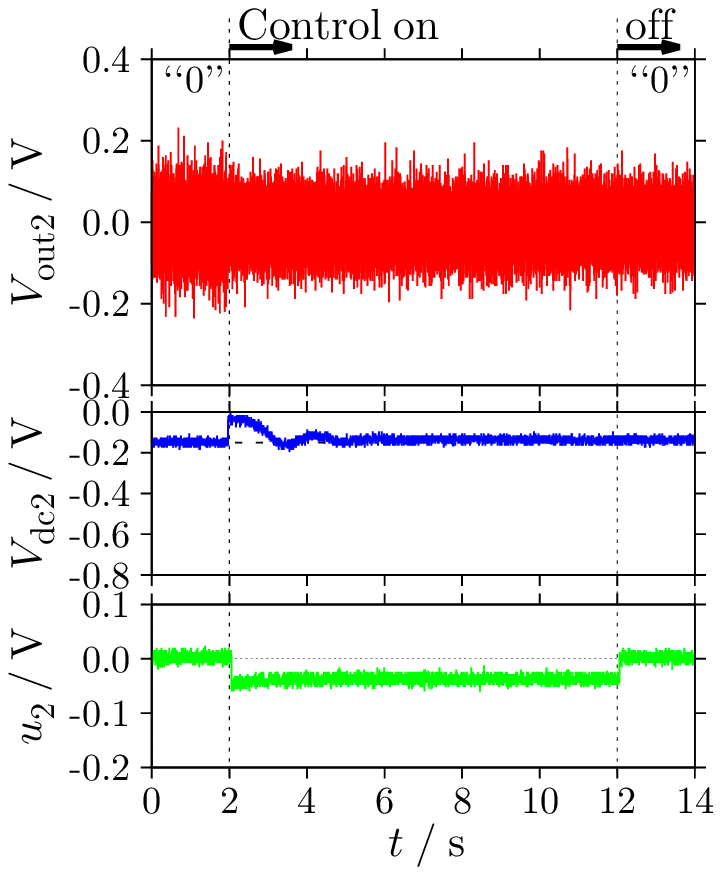}
       \label{00to01Res2}}
   \end{minipage}\\
  \hspace{-1mm} \begin{minipage}{0.47\hsize}
   \subfigure[Switching control from ``1'' to ``0'' in Res.~1 at $V_{\rm ref} =K_{1} V_{\rm ave1}^{{\rm S}2}$.]{
    \centering
    \includegraphics[width=0.99\linewidth]{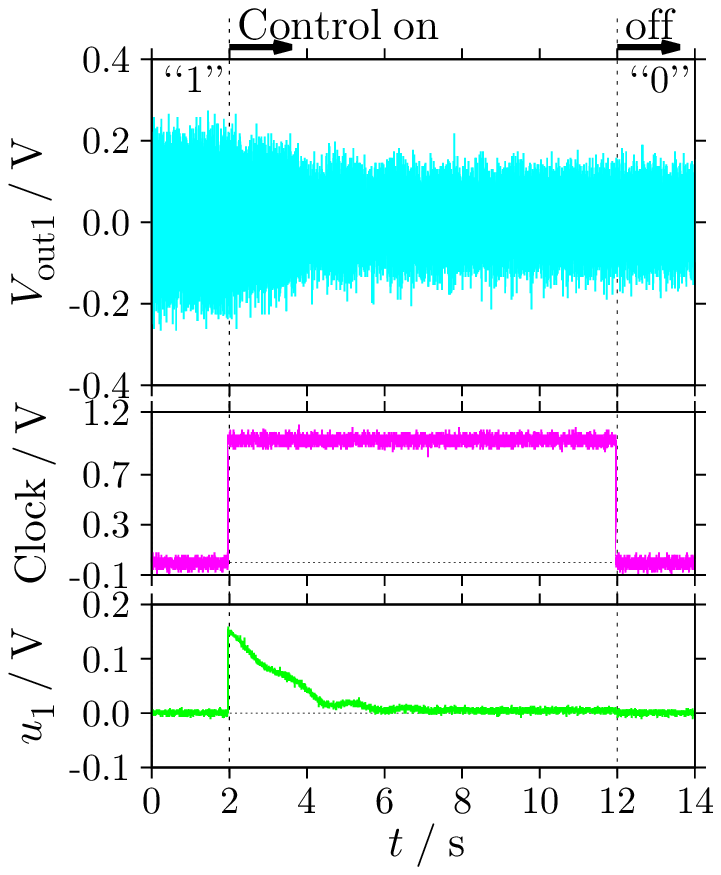}
      \label{01to10Res1}}
  \end{minipage}
  \hspace{1.5mm}
  \begin{minipage}{0.47\hsize}
   \subfigure[Switching control from ``0'' to ``1'' in Res.~2 at $K_{2}=K_{2}^{\rm S}$.]{
    \centering
    \includegraphics[width=0.99\linewidth]{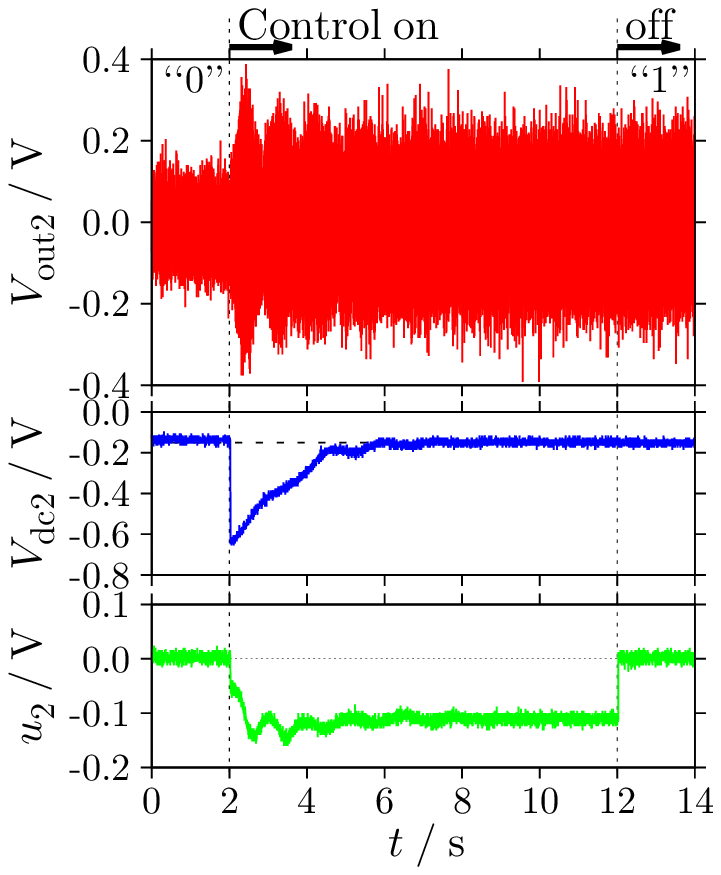}
       \label{01to10Res2}}
   \end{minipage}\\
  \hspace{-1mm}\begin{minipage}{0.47\hsize}
   \subfigure[Switching control from ``0'' to ``1'' in Res.~1 at $V_{\rm ref} = K_{1} V_{\rm ave1}^{{\rm L}2}$.]{
    \centering
    \includegraphics[width=0.99\linewidth]{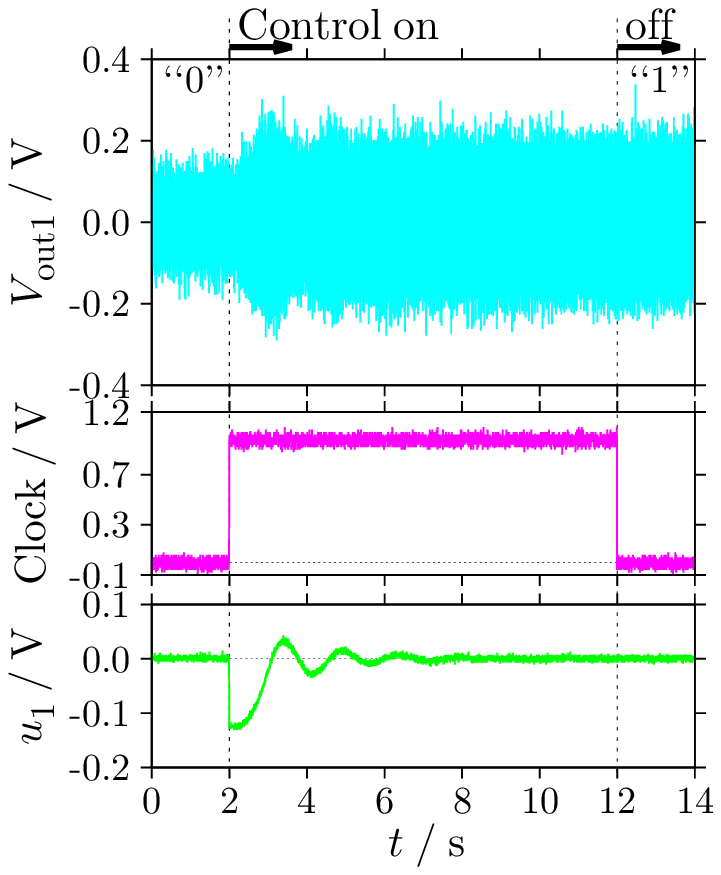}
      \label{10to11Res1}}
  \end{minipage}
  \hspace{1.5mm}
  \begin{minipage}{0.47\hsize}
   \subfigure[Switching control from ``1'' to ``1'' in Res.~2 at $K_{2}=K_{2}^{\rm L}$.]{
    \centering
    \includegraphics[width=0.99\linewidth]{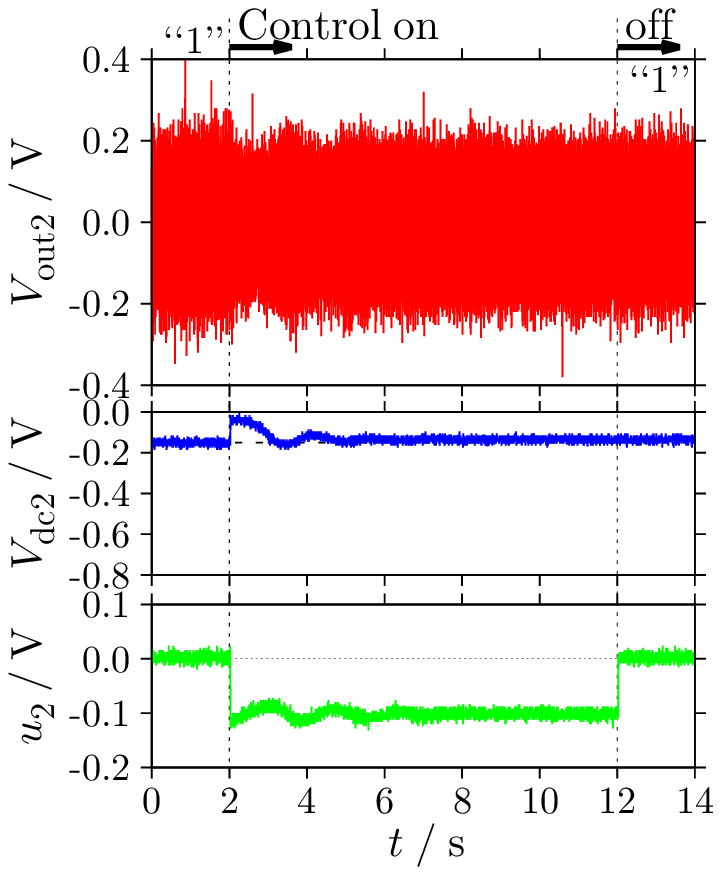}
       \label{10to11Res2}}
   \end{minipage}
 \caption{Switching control in coupled nonlinear MEMS resonator. Aqua and red lines correspond to output voltage $V_{\rm out1}$ in Res.~1 and $V_{\rm out2}$ in Res.~2. Purple, blue, and green lines show clock signal, dc bias voltage $V_{{\rm dc}2}$ of Res.~2, and control input. The control input is applied at $2$\,s from the beginning of the oscillogram as shown in the first vertical dashed line. The second vertical dashed line at $12$\,s represents the moment the control ends. External reference signal $V_{\rm ref}$ and gain $K_{2}$ are switched at $2$\,s.} 
  \label{00to11}
  \end{center}
\end{figure}

In order to realize the 2-bit binary counter as shown in \tabref{count}, 
we apply the proposed switching control to two MEMS resonators. 
\Figref{00to11} shows the experimental results of the switching control. 
\Figrefs{00to01Res1}, (c), and (e) (\Figrefs{00to01Res2}, (d), and (f)) correspond to the switching control results in Res.~1 (Res.~2). 
When the clock signal is set at $0$\,V ($1$\,V), the control input is off (on). 
The external reference signal $V_{\rm ref}$ and the dc bias voltage $V_{{\rm dc}2}$ of Res.~2 are switched by the rising edge of the clock signal. 

The switching control from ``00'' to ``01'' is implemented as shown in \Figrefs{00to01Res1} and (b). 
These results show that Res.~2 stays at ``0'' and Res.~1 changes from ``0'' to ``1''. 
\Figrefs{01to10Res1} and (d) show the switching control from ``01'' to ``10''. 
When Res.~1 changes from ``1'' to ``0'', it triggers the switching control of Res.~2. 
Note that the absolute value of the excitation force (the sum of the dc bias voltage and the control input) in \Figref{01to10Res2} 
exceeds that in \Figref{00to01Res2} when the control input is applied. 
As a result, Res.~2 changes from ``0'' to ``1''. 
\Figrefs{10to11Res1} and (f) show that the present state is ``10'' and the next state becomes ``11''. 
The transition is slow from ``0'' to ``1'' in Res.~1 and Res.~2 stays at ``1''. 
After the switching control was completed, 
the control input $u_{1}$ of Res.~1 disappeared due to the feedback control as shown in \Figrefs{00to01Res1}, (c), and (e). 
It was confirmed that the dc bias voltage $V_{{\rm dc}2}$ of Res.~2 becomes $-0.15$\,V when the state of Res.~1 converges to a steady state. 
We found that the switching control results are related to the initial states in two MEMS resonators. 
Here, when friction and stiction occur in comb-drive resonators,
we could not control desired switching behaviors between coexisting stable states as a binary counter. 

\begin{figure}[!b]
 \centering
 \includegraphics[width=0.6\linewidth]{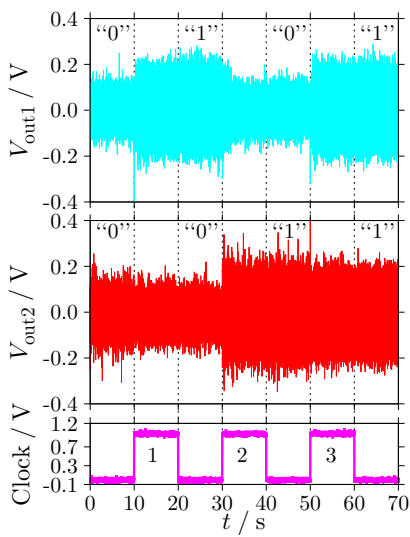}
 \caption{Switching control sequence (``00'' $\rightarrow$ ``01'' $\rightarrow$ ``10'' $\rightarrow$ ``11'') as 2-bit binary counter in electrically coupled nonlinear MEMS resonators.}
\vspace{-0mm}
\label{SwitchingControl}
\end{figure}

\figref{SwitchingControl} shows the oscillogram of the switching control sequence. 
The states in two MEMS resonators start from ``00'' and continues to ``01'', ``10'', and ``11'' at each clock signal. 
Here, the outputs must repeat the binary count sequence with a return to ``00''. 
However, the switching from ``11'' to ``00'' is not realized in the proposed 
switching system. 

In order to implement the switching from ``11'' to ``00'', we propose 
that a reset operation is applied to the coupled MEMS resonators. 
When the control input is not applied, every initial state converges toward either of two stable states (``0'' and ``1'') 
due to two basins of attraction. 
The small amplitude solution has the basin of attraction around the origin~\cite{Unterreithmeier2010b, Atsushi2012}. 
Based on these preceding results, the reset operation can be realized. As a result, we confirmed 
that two coupled MEMS resonators can be used as a $2$-bit binary counter by using the proposed switching control and the reset operation.

%%%%%
%%%%%
\section{Conclusion}\label{section5}
We investigated a binary counter that consists of a coupled system of MEMS resonators with nonlinear characteristics. 
It was confirmed that the switching control results are related to the initial states in coupled resonators. 
To the best of our knowledge, 
we experimentally realized the novel switching control sequence (``00'' $\rightarrow$ ``01'' $\rightarrow$ ``10'' $\rightarrow$ ``11'') 
as a 2-bit binary counter that consists of coupled nonlinear MEMS resonators. 
It was also proposed that a reset operation is applied to MEMS resonators for the switching control from ``11'' to ``00''. 
Since our proposed switching control system is only a prototype, its performance will be improved by an optimized switching control method. 
Nevertheless, this study is the world's first implementation of a logic system consisting of electrically coupled nonlinear MEMS resonators. 
In addition, this paper opens the way for further investigations of the switching control of coexisting states in coupled nonlinear dynamics. 

\section*{Acknowledgments}
This work was supported in part by the Global COE of Kyoto University, Regional Innovation Cluster Program 
``Kyoto Environmental Nanotechnology Cluster'', and JSPS KAKENHI (Grant-in-Aid for Exploratory Research) $\sharp$21656074. 
The authors would like to express our appreciation to Dr.~S.~Naik (SPAWAR, USA) for fruitful discussions and supports to design MEMS resonators. 
One of the authors (AY) would like to thank Mr.~Y.~Tanaka (Department of Electronic Science and Engineering, Kyoto University) 
for comments through discussions. 

\bibliographystyle{elsarticle-num}

\end{document}